\documentclass[10pt,letterpaper, twocolumn]{article} 

\usepackage{ol2}
\usepackage{amsmath}
\usepackage{amssymb}  
\def\ffi {fractional frequency instability}
\def\ffs {fractional frequency stability}

\def\fwhm {full-width at half-maximum}
\def\mw {$\lambda_{m}$}

\def\MOT {magneto-optical trap}

\newcommand{\si}{$\sim$}
\newcommand{\microns}{$\mu$m}
\newcommand{\uK}{$\mu$K}  
\newcommand{\uW}{$\mu$W}

\newcommand{\ee}[1]{\ensuremath{\times 10^{ #1}}}

\newcommand{\clockTboth}{$^{1}S_{0} \leftrightarrow\,^{3}P_{0}$} 
\newcommand{\clockTdash}{$^{1}S_{0} -\,^{3}P_{0}$} 
\newcommand{\st}{$\sigma_{y}(\tau)$}
\newcommand{\Hg}{$^{199}$Hg}

\begin{document}

\twocolumn[ 

\title{Laser locking to the $^{199}$Hg \clockTdash\ clock transition with  5.4\ee{-15}/$\sqrt{\tau}$ fractional frequency instability}


\author{J.~J.~McFerran,$^{1,*}$ D.~V.~Magalh\~{a}es,$^2$ C.~Mandache,$^1$ J. Millo,$^1$ W.~Zhang,$^1$   Y.~Le~Coq,$^1$ G. Santarelli,$^1$ and S.~Bize,$^1$} 

\address{
LNE-SYRTE, Observatoire de Paris, CNRS, UPMC, 61 Avenue de l'Observatoire, 75014 Paris, France
\\
$^2$Escola de Engenharia de S\~{a}o Carlos, Universidade de S\~{a}o Paulo, S\~{a}o Carlos, Brazil \\
}

\begin{abstract}
With \Hg\ atoms confined in an optical lattice trap in the Lamb-Dicke regime, we obtain a spectral line at 265.6\,nm   in which the full-width at half-maximum is $\lesssim$ 15\,Hz.                                                                                                                                                                                                                                                                                                                                                                                         Here we lock an ultrastable laser to this ultranarrow  \clockTdash\  clock transition and achieve a \ffs\ of 5.4\ee{-15}/$\sqrt{\tau}$  for  $\tau\leq400$\,s. 
The highly stable laser light used for the atom probing is derived from a 1062.6\,nm fiber laser locked to an ultrastable optical cavity that exhibits a mean drift rate of  -6.0\ee{-17} s$^{-1}$  (-16.9\,mHz s$^{-1}$ at 282\,THz) over a five month period. 
A comparison between  two such  lasers locked to independent optical cavities shows a flicker noise limited fractional frequency instability of 4\ee{-16} per cavity.
\end{abstract}




 ] 

\noindent 
Optical atomic frequency references are making significant advances in terms of accuracy~\cite{Lud2008, Ros2008, Lem2009a, Cho2010, Hun2012} and stability~\cite{Oat2000,Tak2011, Jia2011} across a range of atomic species, 
which is important for continued investigations into  a potential redefinition of the SI second, and into possible variations of
 fundamental constants~\cite{Kar2008}.
It also motivates research into establishing low-noise frequency links between various atomic clocks over distant Earth locations via optical fibers~\cite{Lop2010, Pre2012}  and space~\cite{Pie2008,Cac2009}.   
In previous work we established the potential of the 
  (6$s^{2}$)  $^{1}S_{0} -$ ($6s6p$) $^{3}P_{0}$ clock  transition in \Hg\ as a high-accuracy atomic frequency reference~\cite{McF2012}.  All the associated frequency measurements to date have relied on line-center determinations from spectral measurements.  Here we lock a probe laser to the \Hg\ clock transition  for the first time  
  and deduce the \ffs, \st: demonstrating a $\tau^{-1/2} $ dependence typical of atomic clocks. 
  With  \st\ integrating  down to  few  times $10^{-16}$ within several hundred seconds, 
  accuracies in the $10^{-17}$ range are foreseeable in forthcoming work. 
   We also present \st\ measurements between two fused-silica mirror based high-finesse optical cavities, one of which is used for the \Hg\ spectroscopy, demonstrating a flicker noise floor of 4\ee{-16} per cavity. 

 Many  components of the \Hg\ experiment  have been described previously~\cite{Pet2008, Mej2011}, so we only highlight a few  elements below.
Atoms are confined in a vertically orientated optical lattice  loaded from a single stage \MOT. 
 Once trapped and with the MOT fields off, we apply a Rabi light pulse  
 at the \clockTboth\ transition frequency.  At present we rely solely on ground state detection. 
There are (2-3)\ee{3} atoms trapped in the lattice from which  a spectral line with \si11\,Hz \fwhm\ can be generated~\cite{McF2012}.   

 The source for the cooling light is a thin-disk Yb:YAG laser that is wavelength tuned using a birefringement filter and a  temperature controlled etalon to 1014.902\,nm. 
This light is frequency quadrupled through two resonant nonlinear optical  cavities, the first producing \si3.5\,W of 507.4\,nm light and the second \si70\,mW of 253.7\,nm UV light.
A challenging aspect of the mercury clock experiment is to produce a sufficiently deep optical lattice trap.  The lower atomic polarizability of Hg in comparison to Sr and Yb implies that at least ten times more optical power is required to produce the same lattice depth.  
The lattice light generation begins with \si 900\,mW of Ti:sapphire laser light that is frequency doubled in a resonant doubling cavity to produce 160\,mW of 362.570\,nm light, near the magic wavelength, \mw.
 The value of \mw\ has previously been determined to within 3\,ppm\,~\cite{McF2012}.
 Two spherical mirrors with radius of curvature equal to 250\,mm   and reflectivity of 99.5\,\% at 360\,nm form the lattice build up cavity. 
 Here we produce 6.5 ($\pm0.7$)\,W of intracavity circulating power. 
  The waist of the cavity with $w_{0} (e^{-2})=120$\,\microns\ coincides with the MOT atom cloud (the atom cloud can be displaced to optimize overlap with the lattice light) and the $e^{-2}$ radius of the MOT atom cloud is \si110\,\microns\  prior to trap loading.   Here we produce a lattice depth of 22 times the associated recoil energy, $E_{\mathrm{R}}$, (\si8\,\uK).

 \begin{figure}[h]
 \begin{center}
{		
  \includegraphics[width=7.0cm,keepaspectratio=true]{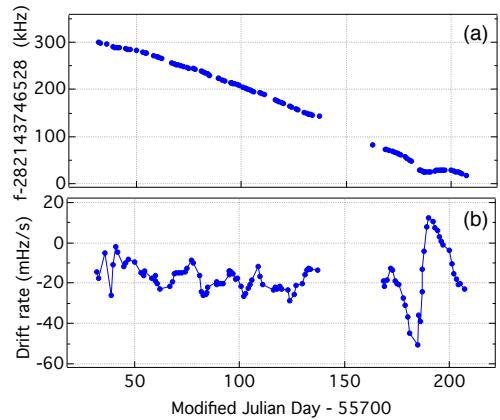}} 
\caption[]{\footnotesize    Frequency and drift-rate record of a fiber laser at 1062.6\,nm locked to an  ultrastable cavity with a 10\,cm length ULE spacer and fused silica mirrors, recorded against the \Hg\ clock transition.   The mean drift rate is -16.9\,mHz s$^{-1}$ (-6.0\ee{-17} s$^{-1}$).
  } \label{USLcavity}	
\end{center}
\end{figure}


 \begin{figure}[h]
 \begin{center}
{		
  \includegraphics[width=7.8cm,keepaspectratio=true]{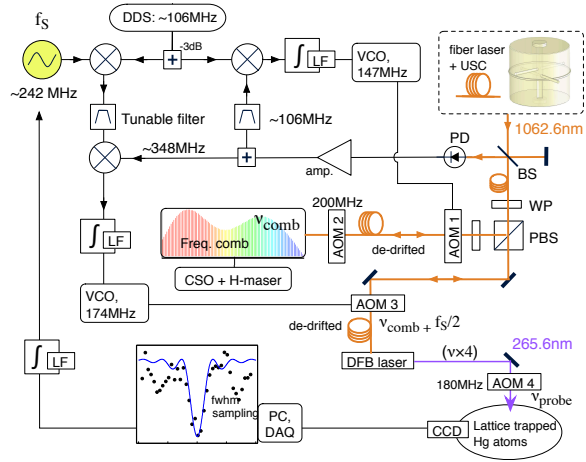}} 
\caption[]{\footnotesize   (Color online) Scheme for locking to the \Hg\ clock line and measuring the line-center frequency.   
A data acquisition system is used to create a correction signal that is delivered to the $f_{\mathrm{S}}$\si242\,MHz synthesizer, which controls  the frequency of the probe light  via AOM 3.  
 AOM, acousto-optic modulator; CSO, cryogenic sapphire oscillator; DAQ, data acquisition and sequence control; DFB, distributed feedback diode laser;  DDS, direct digital synthesizer; 
 LF, loop filter; PBS, polarizing beam splitter; PD, photodiode; USC, ultrastable cavity; VCO, voltage controlled oscillator; WP, wave plate.
  } \label{Hglockschema}		
\end{center}
\end{figure}

The light for probing the clock transition at 265.6\,nm commences with a cavity-stabilized  fiber laser at 1062.6\,nm that is frequency quadrupled through two resonant frequency doubling stages. 
  To provide a probe signal with  sufficiently narrow linewidth and high stability, the fiber laser is locked to an ultrastable optical cavity using Pound-Drever-Hall locking.   Details of the system are presented in \cite{Daw2010}.   In Fig.~\ref{USLcavity}(a) and (b) we show the frequency and  drift rate of the 1062.6\,nm laser locked to the ultrastable cavity over a six month period (20$^{th}$ June to 12$^{th}$ December 2011). The day to day drift rate is determined by comparisons with the \Hg\ clock transition.  
 There are two characteristics of note.  (i) The absolute drift rate rarely rises above 30\,mHz s$^{-1}$; the mean for the data  in Fig.~\ref{USLcavity}(b) is  -16.9\,mHz s$^{-1}$, or in fractional frequency terms -6.0\ee{-17} s$^{-1}$. The excursion seen at the end of the record is due to an abnormal temperature variation in the laboratory. (ii) The drift rate is rarely positive, indicating that the cavity length is slowly lengthening over time, presumably due to creep in the optical cavity spacer (ULE), which is orientated vertically. 
 
  To generate adequate power in the UV for the probe, about 400\,\uW\ of  ultrastable laser (USL) light is used to injection lock a 250\,mW distributed feedback semiconductor laser before carrying out the frequency quadrupling. Information regarding the frequency doubling resonant cavities is presented in~\cite{Daw2010}.
 \begin{figure}[h]
 \begin{center}
{		
  \includegraphics[width=7.8cm,keepaspectratio=true]{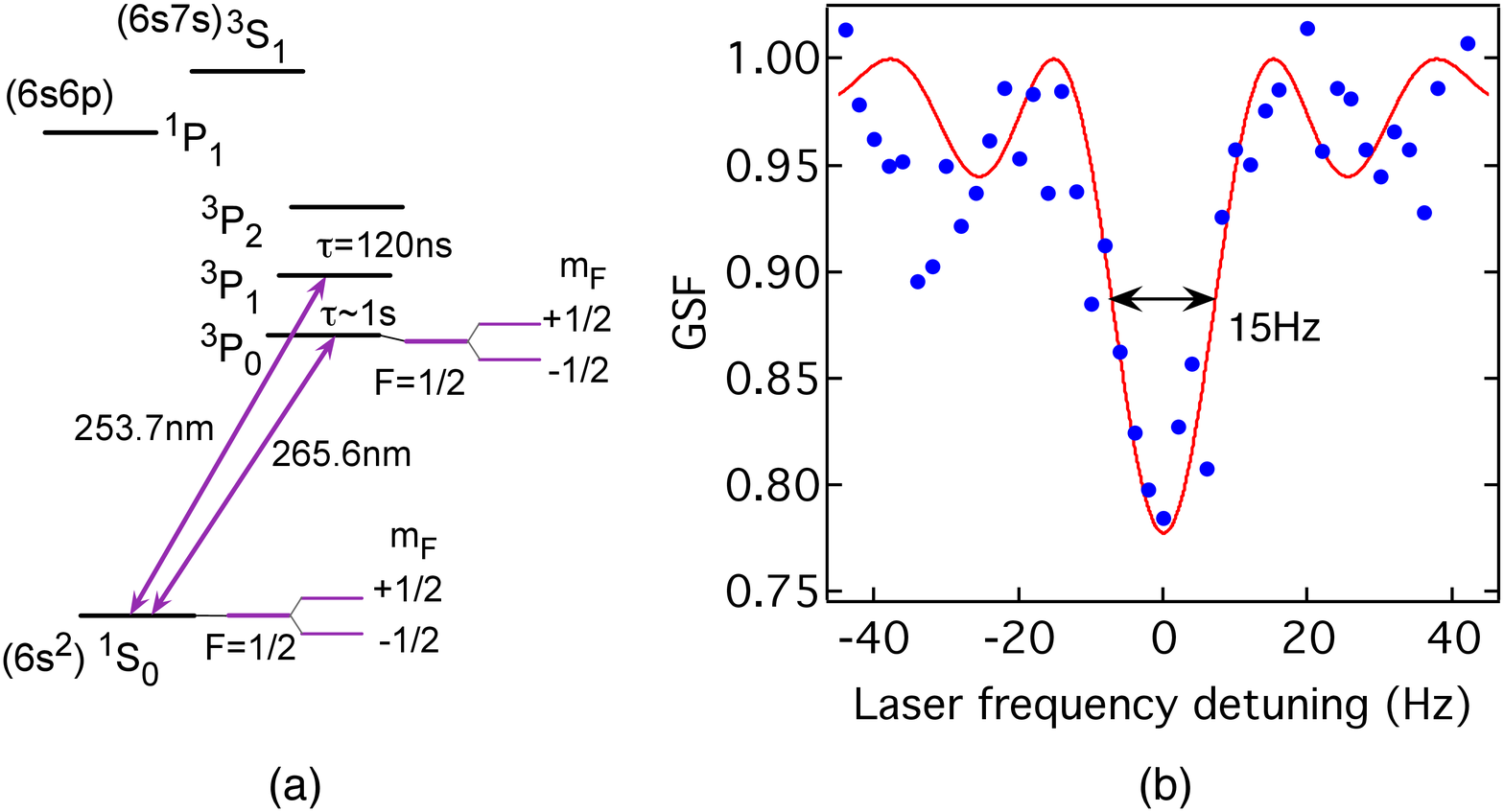}} 
\caption[]{\footnotesize(Color online)  (a) Partial term diagram for \Hg\ showing the cooling (253.7\,nm) and clock (265.6\,nm) transitions. (b) Ground state fraction versus probe laser detuning frequency for a 50\,ms square probe pulse, (solid line) curve fit with $\Omega.T=1.3 \pi$\,rad. 
The Fourier transform limited width of the pulse is 16\,Hz.
  } \label{UltranarrowSpectrum}		
\end{center}
\end{figure}
Despite the very low drift rate exhibited by the USL we find it accommodating to implement a drift cancellation scheme where a digital synthesizer is used to compensate the frequency drift of the probe light interacting with the atoms.   The scheme is illustrated in Fig.~\ref{Hglockschema}.  
Both the light reaching the atoms and the light received by a frequency comb experience the same drift cancellation (with a factor of 4 difference). 
The 1062.6\,nm laser light is delivered to the main Hg table and the frequency comb by way of polarization maintaining optical fiber, for which noise induced by the fiber is actively cancelled.  
The frequency shift produced by AOM 3  
 used in the noise cancellation process also provides a means of sweeping the frequency of the probe light reaching the atoms with the aid of a computer to control a  synthesizer (at $f_{\mathrm{S}}\sim242$\,MHz) in the control path of AOM 3. 
  The computer records  $f_{\mathrm{S}}$, which is used to find the absolute frequency after incorporating the frequency comb measurements of the 1062.6\,nm light and the 
 offset from AOM 4, such that $\nu_{\mathrm{probe}}=4 (\nu_{\mathrm{comb}}+f_{\mathrm{S}}/2)-f_{\mathrm{aom4}}$ (refer to Fig.~\ref{Hglockschema}).  
 The clock and cooling transitions are shown in the partial term diagram of Fig.~\ref{UltranarrowSpectrum}(a).
An example of the clock transition spectrum is seen in Fig.~\ref{UltranarrowSpectrum}(b), obtained with a 50\,ms duration square probe pulse and a cycle time $T_{c}=1.46$\,s. The trace was averaged over $N=4$ scans. The Rabi angle for the line fit is $1.3\pi$ rad.  
From this data, we can estimate the cycle to cycle noise in the measurement of the transition probability to be $\sigma_{\delta P}=0.1$, which is $\sqrt{N}=2$ times larger than apparent in Fig.~\ref{UltranarrowSpectrum}(b) due to the averaging. 
  Using the relationship, 	
\begin{equation}
\sigma_{y}(\tau) =\frac{1}{\nu_{\mathrm{Hg}}}\frac{\sigma_{\delta P}}{\mathrm{d}P/\mathrm{d}\nu}\sqrt{\frac{T_{c}}{\tau}}
\end{equation}
where  $\mathrm{d}P/\mathrm{d}\nu$ is the maximum slope of the resonance,   we can estimate the achievable \ffs, $\sigma_{y}(\tau)$, when locking to the \Hg\  line spectrum.   With  $\mathrm{d}P/\mathrm{d}\nu=0.02$\,Hz$^{-1}$, we find  $\sigma_{y}(\tau) \sim 5.3\times10^{-15}/\sqrt{\tau}$. 

  

\begin{figure}[]
 \begin{center}
{		
  \includegraphics[width=7.2cm,keepaspectratio=true]{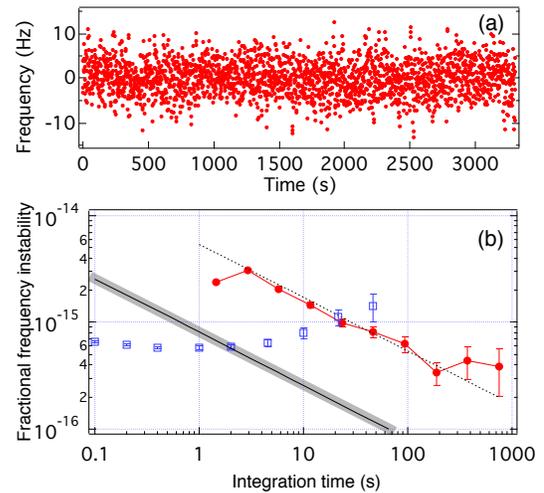}} 
\caption[]{\footnotesize (a)  (Color online) \Hg\ frequency (offset by $\nu_{\mathrm{Hg}}=1 128 575 290 808 162.0$\,Hz) versus time. (b) Fractional frequency instability plot: 
  (filled circles) ultrastable laser locked to the \Hg\  clock transition derived from $f_{\mathrm{S}}$, (open squares) comparison between two ultrastable optical cavities and (solid line) Dick-effect limited instability.  
  } \label{SRAV}		
\end{center}
\end{figure}

By sampling the clock transition at the points of half-maxima (either side of center) we derive a correction signal that is delivered  to AOM 3 (and AOM 1)
 of Fig.~\ref{Hglockschema}, so that the frequency of the probe light becomes locked to the \Hg\ clock transition.
In order to evaluate the clock instability,
 the comb derived IR frequency is averaged and only the long term drift is taken into account (since  the noise of the microwave reference in the comb measurements 
  dominates that of the USL). 
 This way the short term fluctuations of the measured $\nu_{\mathrm{probe}}$ are due solely to that imposed by the atomic resonance.
An example of $\nu_{\mathrm{probe}}$ versus time is shown in Fig.~\ref{SRAV}(a), where the frequency is offset by that reported in \cite{McF2012}.   The cycle time is 1.46\,s, of which the MOT loading time is 1.32\,s.  The \ffi\ for this data (Fig.~\ref{SRAV}(b), filled circles) exhibits $\sigma_{y}(\tau) = 5.4\times10^{-15}/\sqrt{\tau}$ out to 400\,s, in good agreement with the instability predicted above.   

 For comparison, an assessment of \st\ between two USL cavities is shown by the hollow squares of Fig.~\ref{SRAV}(b).   
 One USL is that described above with $\lambda=1062.6$\,nm, while the other  is a similar design, but has the ULE spacer horizontally  orientated~\cite{Mil2009}.  A separate Yb:fiber laser  is locked to each cavity, having a frequency difference of \si1\,GHz (the USLs are in separate laboratories). The minimum fractional instability obtained is 5.7\ee{-16}. Assuming equal contributions from the cavities, each exhibits a \st\ = 4.0\ee{-16} flicker floor.   
We also show the limitation to \st\ set by the Dick-effect for our present probe sequence. The calculation uses an estimation of the USL noise based on the measurement of Fig.~\ref{SRAV}(b) (hollow squares) along with earlier recorded USL vs H-maser data~\cite{Daw2010}.
    The limiting stability is 8.1\ee{-16}$/\sqrt{\tau}$, lying well below that shown for the \Hg\ resonance lock.  These results show that further gains in the \Hg-locked laser frequency stability can be made by improving the S/N of the clock transition.
 At present the contrast of the clock resonance is \si15\,\%.  With a deeper lattice trap (e.g. with a reduced beam waist) and the use of atom number normalization methods we expect to increase the contrast and S/N such that the  very low Dick-effect  instability limitation is reached. 

To conclude, we have demonstrated a fractional frequency instability of 5.4\ee{-15}$/\sqrt{\tau}$ for a 265.6\,nm laser light locked to the clock transition of \Hg, 
 and a short term instability of   5.7\ee{-16} between two lasers locked to separate 10\,cm long cavities.  One optical cavity exhibited a mean drift rate of -6.0\ee{-17}\,s$^{-1}$ over a six month period; one of the lowest reported for optical cavities at RT. 
  We also show the first  laser lock  to  a   spectral line of ultracold neutral atoms in the ultraviolet domain, to our knowledge.
 Based on one day's averaging, the fractional frequency uncertainties of various systematic shifts for the \Hg\ clock should reach into the 10$^{-17}$ range.

The authors thank the  Syst\`{e}mes de R\'{e}f\'{e}rence Temps-Espace technical support, in particular M. Lours and F.
Cornu.  This
work is partly funded by IFRAF and CNES. 
%

\newpage




\end{document}